\documentclass[12pt]{article}
\usepackage{graphicx}
\usepackage{dcolumn}
\usepackage{bm}
\usepackage{graphics,exscale}
\usepackage{amsmath}
\usepackage{amssymb}
\usepackage{amscd}
\usepackage{afterpage}
\usepackage{float,times}
\usepackage{subfigure}
\usepackage{rotating}
\usepackage{multirow}
\usepackage{fancyheadings}
\usepackage{epsfig}
\usepackage{theorem}
\usepackage{moreverb}
\usepackage{euscript}
\usepackage{psfrag}
\newcommand{\rs}{RS$_7$\ }
\begin{document}
\pagestyle{empty}
{\hbox to\hsize{\hfill May 2009 }}

\vspace*{15mm}
\begin{center}

{\Large\bf Warping the Universal Extra Dimensions}\\
\vspace{1.5cm}

{\large Kristian L. McDonald\footnote{Email: klmcd@triumf.ca}}\\
\vspace{1.0cm}
{\it {Theory Group, TRIUMF, 4004 Wesbrook Mall, Vancouver, BC V6T2A3, Canada.}}\\
\vspace{1.4cm}

\end{center}
\begin{abstract}
We develop the necessary ingredients for the construction of realistic models with warped universal extra dimensions. Our investigations are based on the seven dimensional (7D) spacetime $AdS_5\times T^2/Z_2$ and we derive the Kaluza-Klein (KK) spectra for gravitons, bulk vectors and the TeV brane localized Higgs boson. We show that, starting with a massive 7D fermion, one may obtain a single chiral massless mode whose profile is readily localized towards the Planck or TeV brane. This allows one to place the standard model fermions in the bulk and construct models of flavor as in Randall-Sundrum models. Our solution also admits the familiar KK parity of UED models so that the lightest odd KK state is stable and may be a dark matter (DM) candidate. As an additional feature the $AdS_5$ warping ensures that the excited modes on the torus, including the DM candidate, appear at TeV energies (as is usually assumed in UED models) even though the Planck scale sets the dimensions for the torus.
\end{abstract}

\vfill

\eject
\pagestyle{empty}
\setcounter{page}{1}
\setcounter{footnote}{0}
\pagestyle{plain}
\section{Introduction}
The last decade has seen much interest in the possibility that nature may possess phenomenologically relevant extra spatial dimensions. Much of this work has been spawned by the hierarchy problem of the standard model (SM), with the study of large extra dimensions permitting one to phrase the hierarchy problem in a new way~\cite{ArkaniHamed:1998rs}. Alternatively a warped space allows one to realize the weak scale as a red shifted incarnation of Planck scale sized input parameters~\cite{Randall:1999ee}. Extra dimensions also provide new tools for constructing models of flavor, with bulk fermions permitting the fermion mass hierarchies to emerge from higher dimensional wavefunction overlaps~\cite{ArkaniHamed:1999dc}. This idea finds a natural home in the Randall-Sundrum (RS) framework where bulk five dimensional fermions possess Dirac masses whose values control the localization of SM fermions in the warped space~\cite{Grossman:1999ra,Gherghetta:2000qt}. The localization of lighter (heavier) SM fermions towards the Planck (TeV) brane provides a natural suppression (relative enhancement) of their coupling to the SM Higgs and admits the construction of realistic flavor structures~\cite{rs_flavor}. As RS models admit both a solution to the hierarchy problem and a theory of flavor they provide an appealing candidate for the physics beyond the SM.

Besides the hierarchy problem and flavor puzzle of the SM there exist other aspects of nature which extra dimensions may help us to understand. For example, we currently do not understand the repeating structure manifested by the three observed generations of fermions. Even in grand unified theories the three generations are typically built in by hand. Furthermore if the ultraviolet cutoff of the SM is relatively low, say 10 TeV or so, the SM does not explain the stability of the proton, given that baryon number is an accidental symmetry. There is also a dearth of evidence suggesting the existence of dark matter (DM) in the universe, for which the SM has no compelling candidate. 

Interestingly if all of the SM fields propagate in flat extra dimensions, as in models with universal extra dimensions (UED)~\cite{Appelquist:2000nn}, candidate solutions to the above puzzles reveal themselves. If the SM fields propagate in two additional dimensions compactified on a torus then: (a) anomaly constraints require that the number of generations $n_g$ obeys $n_g=0$~mod~3~\cite{Dobrescu:2001ae}, providing a compelling reason for the existence of three generations and (b) a remnant discrete subgroup of the higher dimensional Lorentz group can forbid the most dangerous proton decay operators~\cite{Appelquist:2001mj}. Thus consideration of the spacetime $\mathcal{M}_4\times T^2$ appears to be well motivated\footnote{The torus should be orbifolded as either $T^2/Z_2$~\cite{Appelquist:2000nn} or $T^2/Z_4$ (the chiral square)~\cite{Dobrescu:2004zi} to produce chiral four dimensional fermions. We note that a recent work has considered a UED model with the extra space $S^2/Z_2$~\cite{Maru:2009wu}.}. 

Another appealing feature of UED models is the existence of a Kaluza-Klein (KK) parity, which forces the lightest parity odd KK particle (LKP) to be stable and thus a DM candidate~\cite{Servant:2002aq} (for a review see~\cite{Hooper:2007qk}). The LKP makes an attractive DM candidate because, as far as extensions of the SM are concerned, UED models are quite predictive. In UED models the field content of the SM is not enlarged (modulo a right-chiral neutrino), it is simply placed in a higher dimensional spacetime, so the interactions of the KK states are governed by known SM coupling constants and the parameters of the SM Higgs potential.

Whilst having a number of interesting features, UED models do not shed any light on the flavor puzzle\footnote{Minimal UED models have a trivial flavor structure with fermion mass hierarchies emerging from Yukawa hierarchies put in by hand. Non-minimal UED models, like Split-UED~\cite{Park:2009cs}, may admit flavor structures~\cite{J_Shu}.} and do not provide a solution to the hierarchy problem. The RS model is not a UED model because the Higgs boson typically resides on the TeV brane to realize the weak scale via warping. However even if the Higgs propagates in the bulk, as in the holographic composite Higgs models~\cite{Contino:2003ve} and models with an ``off the wall Higgs''~\cite{Davoudiasl:2005uu}, the warping necessarily breaks the translational invariance along the extra dimension so that no KK parity remains. Thus a DM candidate typically has to be added to RS models as an additional ingredient~\cite{Panico:2008bx}, although it is possible to construct a KK parity in RS models by gluing together multiple warped throats and imposing a discrete interchange symmetry on the throats~\cite{Agashe:2007jb}. 

It is the goal of the present work to combine the appealing features of RS models, namely a theory of flavor and a warped realization of the weak scale, and UED models. We are primarily interested in realizing a warped space model with a KK parity to ensure the presence of a stable LKP. In such a model the underlying geometry of spacetime would be responsible for the suppression of the weak scale, the hierarchy of observed fermion masses and the existence of a stable DM candidate. We find the possibility that such diverse puzzles may have a common solution in terms of an underlying geometrical structure quite interesting. Such a framework would possess warped universal extra dimensions (WUED) and we develop some of the necessary ingredients for the construction of such models in this work.

It would seem that the simplest extension of the RS model that incorporates the UED KK parity would be to consider the spacetime $AdS_5\times S^1$ (referred to as RS6 in~\cite{Davoudiasl:2002wz,Davoudiasl:2008qm}) with all SM fields propagating on the circle. However in 6D the minimal fermion is a chiral spinor with four non-zero components and 6D chirality precludes a bulk mass for such a fermion. Thus even if one could obtain a chiral massless mode in the effective 4D theory one would not retain the RS approach to flavor (which utilizes bulk fermion mass parameters to produce independent bulk wavefunctions for distinct SM fields). 

In seven dimensions a bulk fermion is vectorial and may possess a bulk mass so that, in principle, the RS mechanism of flavor may be viable. There are multiple ways to extend the RS model to a 7D spacetime. If one considers the warping to occur on a slice of $AdS_5$, and not $AdS_{6}$~\cite{Gherghetta:2000qi} or $AdS_7$~\cite{Appelquist:2002ft,Bao:2005ni}, then two possibilities present themselves; namely $AdS_5\times S^2$ and $AdS_5\times T^2$. In this work we consider the spacetime $AdS_5\times T^2/Z_2$, which we refer\footnote{We reserve RS7 for $AdS_5\times S^2$ as defined in~\cite{Davoudiasl:2008qm}. More generally we may label $AdS_5\times S^{d-5}$ as RS$d$ and $AdS_5\times T^{d-5}$ as RS$_d$, with RS6=RS$_6$. We discuss the gravitational background for the generalized RS$_d$ in Appendix~\ref{app:rsd}.} to as \rs.

The construction of a viable WUED model requires all SM fields to propagate on the torus. As we aim to retain the RS approach to flavor we must consider a bulk 7D fermion to determine if a viable massless chiral mode obtains. A 7D fermion is an eight component object and thus the RS results do not translate \emph{a priori}. Our work shows that, starting with a massive bulk \rs fermion, one indeed obtains a single massless chiral mode. Importantly the localization of this chiral mode is determined by the bulk mass parameter with Planck or TeV brane localization easily achieved. Thus one may model a SM fermion as the zero mode of a bulk \rs fermion and consider the RS approach to flavor. 

In this work we also derive the KK spectra for gravitons, bulk vectors and the TeV brane localized Higgs. We show that the modes corresponding to excitations on the torus generically have order TeV masses even though the dimensions of the torus are set by the Planck scale. This is particularly interesting because the LKP is necessarily a mode which is excited on the torus. Thus the \rs warping motivates order TeV scale KK states, as is usually assumed in UED models, and in particular motivates a stable, order TeV DM candidate for RS models.

As we will show, the KK spectrum of \rs contains the usual RS KK states, with the SM fields possessing the same profiles along $AdS_5$ as they do in RS. The spectra also contains UED-like KK states corresponding to the $T^2$ excitations. The KK parity operates only on the UED-like modes and constrains their interactions. These modes cannot be singly produced by the annihilation of SM fields and the bounds on these states are expected to be weakened as in UED models. As the KK parity does not operate on the RS KK modes these are subject to the usual bounds. Thus the model may be viewed as essentially an RS model with additional order TeV KK states, including a DM candidate, which are constrained by KK parity.

Though our interests are mainly phenomenological, we are partly motivated by the fact that in a generic string theoretic ultraviolet completion of the RS model additional compact spatial dimensions are expected. As discussed in~\cite{Bao:2005ni} one may obtain $AdS_5$ with a stack of parallel $D3$ branes in type-IIB string theory~\cite{Aharony:1999ti}, though additional compact dimensions will be present (see~\cite{Gherghetta:2006yq} for some string-motivated realizations of RS-like scenarios). From a ``bottom up'' perspective it is important to know how the presence of extra dimensions may modify our understanding of the RS model and what new phenomenologically relevant features may emerge.

Works based on 7D warped spaces exist already in the literature; see for example~\cite{Gherghetta:2000jf,Bao:2005ni}. It is known, for example, that in $AdS_7$ the cancellation of boundary anomalies~\cite{Gherghetta:2002xf} necessarily constrains the boundary symmetries and field content. The combination of warped and universal extra dimensions has also been previously considered on a slice of $AdS_7$~\cite{Appelquist:2002ft}. In that work a massless, bulk, gauge-singlet fermion was considered in addition to the 6D TeV brane localized SM fields. This enabled the authors to retain a KK parity on the brane but did not permit the RS approach to flavor. Our approach enables one to peel the SM fields off the brane, thereby retaining both KK parity and RS flavor.

We are interested primarily in retaining RS flavor, however for completeness we briefly discuss the trivial flavor scenario with all SM fields residing on the TeV brane (see Section~\ref{sec:sm_tev}). In that case the model is essentially a UED model with additional KK graviton states. Of interest is the fact that when the SM is localized on the TeV brane in \rs both the weak scale and the UED KK scale are red shifted to the TeV scale. This observation motivates the connection between the weak scale and the UED KK scale usually assumed in the UED literature. It also holds more generally when the UED Lagrangian is localized on the TeV brane in spacetimes of the form $AdS_5\times \mathcal{M}^\delta$, where $\mathcal{M}^\delta$ is a compact space. This differs from the $AdS_7$ case where the KK modes of TeV brane localized fields are at the TeV scale \emph{only} if the the radius of the flat directions is $R^{-1}\sim $~TeV~\cite{Appelquist:2002ft}.  

The layout of this work is as follows. We detail the \rs background and derive the KK spectrum for gravitons in Section~\ref{sec:rs7_grav}. In Section~\ref{sec:bosons} we consider the bosonic sector of the model, namely the TeV brane localized Higgs and the KK spectrum for bulk vectors. Bulk fermions are discussed in Section~\ref{sec:rs7_kk_fermi} and we consider KK parity and the LKP in Section~\ref{sec:dm_kk}. We discuss a UED scenario with all SM fields localized on the TeV brane in Section~\ref{sec:sm_tev} and conclude in Section~\ref{sec:conc}. In Appendix~\ref{app:rsd} we generalize the \rs background and we present some general properties of 7D fermions in Appendix~\ref{app:ferm_not}. The KK spectrum of bulk \rs fermions in some limiting cases is given in Appendix~\ref{app:fermi_kk}.
\section{Gravity on $\mathbf{AdS_5\times T^2}$\label{sec:rs7_grav}}
We consider the metric defined by the seven dimensional spacetime interval
\begin{eqnarray}
ds^2&=&e^{-2\sigma(y)}\eta_{\mu\nu}dx^\mu dx^\nu -\delta_{ab}dy^ady^b-(dy)^{2},\nonumber\\
&\equiv&G_{MN}dx^Mdx^N\label{rs7_metric},
\end{eqnarray}
where $M,N=0,1,2,3,5,6,7$ label the full 7D space, $\mu,\nu=0,1,2,3$ label the 4D subspace and the extra dimensions are labeled by $y^{a}$, $a=1,2$, and $x^7=y$ (the latter being the warped direction). The extra dimensions are compact with $y^{a}\in[-\pi R,\pi R]$, $y\in[-\pi r_c,\pi r_c]$  and the points $y^{a}=\pm\pi R$ ($y=\pm\pi r_c$) identified. For simplicity we take equal radii in the $y^a$ directions and the extra dimensions are orbifolded so that the compact space is
\begin{eqnarray}
(T^2/Z_2')\times (S^1/Z_2),\label{rs7_obc}
\end{eqnarray}
where the action of the orbifold symmetries is defined as
\begin{eqnarray}
Z_2'\ :\  y^a\rightarrow - y^a\quad,\quad Z_2\ :\  y\rightarrow - y.
\end{eqnarray}
A bulk field in the above background is in general specified by two parities $(Z_2',Z_2)=(P',P)$ where $P',P=\pm $. A brane field localized at some slice in the warped direction $y$ on the other hand possesses only the $P'$ parity. We note that the orbifolding (\ref{rs7_obc}) ensures there are no massless gravi-vectors.

The sources realizing the above geometry are a bulk cosmological tensor
\begin{eqnarray}
\Lambda^M_N=\mathrm{diag}(\Lambda,\Lambda,\Lambda,\Lambda,\bar{\Lambda},\bar{\Lambda},\Lambda)\label{rs7_cc},
\end{eqnarray}
 and two codimension one branes,
\begin{eqnarray}
T^{\bar{M}}_{h,v\bar{N}}=\mathrm{diag}(V_{h,v},V_{h,v},V_{h,v},V_{h,v},\bar{V}_{h,v},\bar{V}_{h,v}),\label{rs7_tension}
\end{eqnarray}
localized at $y=0$ and $y=\pi r_c$ respectively. Here $\bar{M},\bar{N}=0,1,2,3,5,6$ are the brane Lorentz indices. The Einstein equations are
\begin{eqnarray}
& &\sqrt{G}\left[R_{MN}-\frac{1}{2}G_{MN}R^{(7)}\right]=\nonumber\\
&&-\frac{1}{4M_*^{5}}\left[\sqrt{G}G_{MP}\Lambda^P_N+\delta^{\bar{M}}_M\delta^{\bar{N}}_N\sqrt{\bar{G}}\bar{G}_{\bar{M}\bar{P}}\left\{T^{\bar{P}}_{h\bar{N}}\delta(y)+T^{\bar{P}}_{v\bar{N}}\delta(y-\pi r_c)\right\}\right],
\end{eqnarray}
where $M_*$ ($R^{(7)}$) is the 7D Planck scale (Ricci scalar), $\bar{G}_{\bar{M}\bar{N}}$ denotes the induced metric at the brane locations and $G=|\mathrm{det}(G_{MN})|$ (similarly for $\bar{G}$). The $(M,N)=(\mu,\nu),(a,a)$ and $(7,7)$ Einstein equations are respectively 
\begin{eqnarray}
2\sigma'^2-\sigma''&=&-\frac{1}{12M^{5}_*}\left\{ \Lambda+V_h\delta(y)+V_v\delta(y-\pi r_c)\right\},\\
5\sigma'^2-2\sigma''&=&-\frac{1}{8M^{5}_*}\left\{\bar{\Lambda}+\bar{V}_h\delta(y)+\bar{V}_v\delta(y-\pi r_c)\right\},\\
6\sigma'^2&=&-\frac{1}{4M_*^{5}}\Lambda,
\end{eqnarray}
with the solution for $\sigma(y)$ being
\begin{eqnarray}
\sigma&=&\sqrt{\frac{-\Lambda}{24M_*^{5}}}|y|\equiv k|y|.\label{sigma_sol}
\end{eqnarray}
The solution requires
\begin{eqnarray}
V_h=-V_v=24kM^{5}_*\quad,\quad V_{h,v}=\frac{3}{4}\bar{V}_{h,v}\quad,\quad\Lambda=\frac{3}{5}\bar{\Lambda},
\end{eqnarray}
whilst the 4D Planck mass is given by
\begin{eqnarray}
M_{Pl}^2= (2\pi R)^2\frac{M^{5}_*}{k}[1-e^{-2k\pi r_c}].\label{rs7_planck}
\end{eqnarray}
As we shall discuss in Section~\ref{sec:dm_kk}, one requires $1\lesssim kR<10$ to keep the tree level mass of the KK partners of the Higgs above $\sim10^2$~GeV. We shall not be considering large values of $R$ so that the fundamental gravity scale is not significantly volume suppressed relative to the Planck scale. We note that the solution (\ref{sigma_sol}) requires $\Lambda<0$ and we define a new  variable by $kz=e^{-ky}$ to rewrite (\ref{rs7_metric}) as
\begin{eqnarray}
ds^2&=&\frac{1}{(kz)^2}\left[\eta_{\mu\nu}dx^\mu dx^\nu-dz^2\right] -\delta_{ab}dy^ady^b.\label{ads_conformal_metric}
\end{eqnarray}
We shall refer to the above setup as $RS_7$ and note that the solution is readily generalized to RS$_d$ with $d>7$ (see Appendix~\ref{app:rsd}).
\subsection{Graviton KK tower\label{sec:rs7_kk_grav}}
To determine the masses and wavefunctions for the KK gravitons one makes the replacement $G_{\mu\nu}=e^{-2\sigma}\eta_{\mu\nu}\rightarrow e^{-2\sigma}(\eta_{\mu\nu}+\kappa h_{\mu\nu})$, with $\kappa=2M_*^{-5/2}$, in the metric (\ref{ads_conformal_metric}). The KK expansion for $h_{\mu\nu}$ is
\begin{eqnarray}
h_{\mu\nu}(x^\tau,y^a,z)=\sum_{\vec{n}}h^{(\vec{n})}_{\mu\nu}(x^\tau)f^{(\vec{n})}_h(z)g^{(n_a)}_+(y^a),
\end{eqnarray}
where\footnote{We emphasize that $n$ ($n_a$) is the quantum number for the warped (flat) direction(s). We shall on occasion also denote $f^{(\vec{n})}_h$ as $f^{(n,n_a)}_h$.} $\vec{n}=(n,n_a)=(n,n_1,n_2)$ and $g^{(n_a)}_+$ are the usual even parity UED wavefunctions which are given in equation (\ref{ued_+_profile}) below. They satisfy $\sum_a\partial_a^2g^{(n_a)}_+= -m_{n_a}^2g^{(n_a)}_+$ where
\begin{eqnarray}
m_{n_a}=\frac{\sqrt{n_1^2+n_2^2}}{R},\label{torus_kk_mass}
\end{eqnarray}
so that $m_{n_a}\sim R^{-1}$ for $n_a\ne 0$. One finds that $f^{(\vec{n})}_h$ must satisfy
\begin{eqnarray}
\left[z^2\partial_z^2-3z\partial_n-\frac{m_{n_a}^2}{k^2}+m_{h,\vec{n}}^2z^2\right]f^{(\vec{n})}_h=0,\label{grav_eom_f}
\end{eqnarray}
where we use the gauge $\partial^\mu h_{\mu\nu}=h^\mu_\mu=0$ and define the KK masses as $m_{h,\vec{n}}$. The profiles obey the orthogonality conditions
\begin{eqnarray}
\int  \frac{dz}{(kz)^{3}}f^{(m,n_a)}_hf^{(m,n_a)}_h&=&\delta^{mn},
\end{eqnarray}
and are given by
\begin{eqnarray}
f_h^{(\vec{n})}(z)=\frac{(kz)^2}{N_h^{(\vec{n})}}\left\{J_{\nu_{ha}}(m_{h,\vec{n}}z)+\beta_h^{(\vec{n})}Y_{\nu_{ha}}(m_{h,\vec{n}}z)\right\},\label{grav_profile}
\end{eqnarray}
where $N_h^{(\vec{n})}$ is a normalization constant and the order of the Bessel functions is
\begin{eqnarray}
\nu_{ha}=\sqrt{4+\frac{m_{n_a}^2}{k^2}}.
\end{eqnarray}
The constants $\beta_h^{(\vec{n})}$, determined by the boundary conditions\footnote{Here $z_*$ denotes the $Z_2$ fixed points, $z_*=z_{0,L}=1/k,e^{k\pi r_c}/k$} $\partial_z f_h^{(\vec{n})}|_{z_*}=0$, are
\begin{eqnarray}
\beta^{(\vec{n})}_h(z_*)=-\frac{J_{\nu_{ha}-1}(m_{h,\vec{n}}z_*)}{Y_{\nu_{ha}-1}(m_{h,\vec{n}}z_*)},
\end{eqnarray}
and the KK masses $m_{h,\vec{n}}$ obtain by enforcing $\beta^{(\vec{n})}_h(z_0)=\beta^{(\vec{n})}_h(z_L)$. The usual 4D graviton has $\vec{n}=0$ and is the only massless mode in the spectrum; all towers with $n_a>0$ do not possess a massless mode. The KK tower $(n,n_a)=(n,0)$ is the usual RS KK tower so that the wavefunctions (\ref{grav_profile}) and the KK masses are identical to the usual RS ones~\cite{Davoudiasl:1999jd}. The wavefunctions and  masses of the modes with $n_2=0$ are also the same as the RS6 gravitons studied in~\cite{Davoudiasl:2002wz}. However in the present work the coupling of these modes to matter and the associated phenomenology will differ to that found in~\cite{Davoudiasl:2002wz}. This is because the WUED framework admits a KK parity which prevents the KK odd states from being singly produced by the annihilation of SM particles. We discuss the KK parity in more detail in Section~\ref{sec:dm_kk} below. 

Let us comment further on the KK masses. As in the RS model, the $n_a=0$ states have order TeV masses despite the fact that $r_c^{-1}\gg$~TeV. The same is true for the $n_a\ne0$ modes; for $R^{-1}\sim k$ these states also appear at TeV energies. This observation is important for the WUED picture as the mass of the LKP will be $\sim$~TeV and not the Planck scale, even if the dimensions of the torus are of order $k^{-1}$. The same is true more generally for the KK modes corresponding to excitations on $T^2$.
\section{Standard Model Bosons on $\mathbf{AdS_5\times T^2}$\label{sec:bosons}}
\subsection{TeV Brane Localized Higgs\label{sec:rs7_higgs}}
The Higgs boson should be localized on the TeV brane in order to realize the weak scale via warping as $\sim e^{-k\pi r_c}M_*$. To ensure the desired UED features the Higgs must propagate in the $y^a$ directions. Thus we consider the SM Higgs as a 6D field localized at $z_L=e^{k\pi r_c}/k$ with action 
\begin{eqnarray}
S_H&=&\int d^{7}x\sqrt{\bar{G}}\left\{G^{\bar{M}\bar{N}}D_{\bar{M}}H^\dagger D_{\bar{N}}H -\lambda(H^2-v_0^2)^2\right\}\delta(z-z_L),
\end{eqnarray}
where $\lambda$ and $v_0$ have dimension $[\lambda]=-2$ and $[v_0]=[H]=2$ respectively. The orbifold action on $H$ is specified by the $Z_2'$ parity which must be $P'=+$ to ensure a zero mode. We KK expand $H$ as
\begin{eqnarray}
H(x^\mu,y^a)=\sum_{n_a}H^{(n_a)}(x^\mu)g^{(n_a)}_+(y^a),
\end{eqnarray}
where $n_a$ denotes the set of integers $n_a=(n_1,n_2)$. The KK masses prior to electroweak symmetry breaking are
\begin{eqnarray}
M_{n_a}^2=-2\lambda v_0^2e^{-2k\pi r_c}+m_{n_a}^2e^{-2k\pi r_c},\label{higgs_kk}
\end{eqnarray}
and we note that provided the relation $R^{-2}> 2\lambda v_0^2$ is satisfied only the $n_a=0$ mode is tachyonic. Also for $n_a\ne0$ the orthogonality relations prevent the presence of a term containing only one $H^{(n_a)}$ field in the potential. Thus there exists a minimum of the scalar potential such that the neutral component of $ H^{(0)}$ develops a non-zero VEV, whilst $\langle H^{(n_a)}\rangle=0$ for all $n_a\ne0$. This permits electroweak symmetry breaking with $\langle H^{(0)}\rangle\sim e^{-k \pi r_c}v_0$ realizing the weak scale via warping as in the RS model.

After electroweak symmetry breaking the boundary Higgs Lagrangian leads to the following KK tower for the Higgs boson
\begin{eqnarray}
m_{h^0,n_a}^2&=&m_{h^0}^2 +m_{n_a}^2e^{-2k\pi r_c},\label{sm_higgs_kk}
\end{eqnarray}
where $m_{h^0}\sim 10^2$~GeV is the SM Higgs mass. The SM Goldstone modes also have KK towers,
\begin{eqnarray}
m_{A^0,n_a}&=& m_{H^{\pm},n_a}=m_{n_a}e^{-k\pi r_c},\label{higgs_goldstone_kk}
\end{eqnarray}
where the $n_a=0$ modes in (\ref{higgs_goldstone_kk}) are the usual Goldstone modes of the SM. As is obvious from (\ref{sm_higgs_kk}) and (\ref{higgs_goldstone_kk}), both the bare mass parameters $v_0,\lambda$ and the KK scale are warped down to order TeV energies.
\subsection{Bulk Vectors on $\mathbf{AdS_5\times T^2}$\label{sec:rs7_kk_vec}}
In this section we consider a bulk $U(1)$ gauge field in the $AdS_{5}\times T^2/Z_2'$ background. As our purpose is to consider the viability of modeling a SM gauge boson by such a state we wish for the vector to have a zero mode. The $(Z_2',Z_2)$ parities of such a bulk gauge field (denoted as $A_M$) are\footnote{We label the $T^2$ polarizations of the bulk gauge boson as $A_a=A_{5,6}$}
\begin{eqnarray}
A_\mu&:&(+,+),\nonumber\\
A_{a,z}&:&(\mp,\pm),\label{rs7_gs_parity}
\end{eqnarray} 
and we note that once one specifies the parities for $A_\mu$ the relative parities for $A_{a,z}$ are fixed. The action for $A_M$ is
\begin{eqnarray}
S_{A}&=&-\frac{1}{4}\int d^{7}x\sqrt{G}\left\{G^{MP}G^{NQ}F_{MN}F_{PQ}\right\},\label{rs7_vec_action}
\end{eqnarray}
and we work with the coordinates defined by (\ref{ads_conformal_metric}). The action (\ref{rs7_vec_action}) leads to an effective 4D action describing a KK tower of vectors and two KK towers of gauge-scalars. The remaining tower of gauge-scalars act as Goldstone modes for the massive vectors. The orbifold parities (\ref{rs7_gs_parity}) were chosen to ensure a massless vector mode, however they also prevent any massless gauge-scalars from appearing in the spectrum. Beyond some passing comments we do not discuss the gauge-scalars in this work. The determination of their KK spectra and the study of their coupling to SM fields are however important issues in this framework.

The action (\ref{rs7_vec_action}) contains mixing terms between the vector modes and the gauge-scalars which may be decoupled by introducing the following bulk gauge fixing term
\begin{eqnarray}
S^{Bulk}_{GF}=-\frac{1}{2\xi}\int d^{7}x\frac{1}{kz}\left(\eta^{\nu\tau}\partial_\nu A_\tau+\xi (kz)[\partial_{z}(K_zA_{z})+\sum_{a}\partial_{a}(K_aA_{a})]\right)^2
,
\end{eqnarray}
where the quantities $K_{a,z}=K_{a,z}(z)$ are defined by
\begin{eqnarray}
K_a\eta^{\nu\tau}=\sqrt{G}G^{aa}G^{\nu\tau}\quad,\quad K_z\eta^{\nu\tau}=\sqrt{G}G^{zz}G^{\nu\tau}.
\end{eqnarray}
Variation of the action $S_A+S^{Bulk}_{GF}$ gives the bulk equations of motion,
\begin{eqnarray}
\sqrt{-G}G^{\mu\tau}G^{\nu\sigma}\partial_\mu F_{\tau\sigma}+\sum_{\bar{a}}\eta^{\mu\nu}\partial_{\bar{a}}[K_{\bar{a}}\partial_{\bar{a}}A_\mu]+\frac{1}{\xi}\frac{1}{kz}\eta^{\mu\tau}\eta^{\nu\sigma}\partial_\mu\partial_\sigma A_\tau&=&0,\label{rs7_eom_vector}\\
\eta^{\mu\tau}\partial_\tau\partial_\mu A_{a}+\xi \partial_{a} [(kz)\sum_{\bar{b}}\partial_{\bar{b}}(K_{\bar{b}}A_{\bar{b}})]-\frac{1}{K_a}\sum_{b}\partial_{b}[\sqrt{G}G^{aa}G^{bb}F_{ab}]&=&0\label{scalar_eom_1},\\
\eta^{\mu\tau}\partial_\tau\partial_\mu A_z+\xi \partial_{z} [(kz)\sum_{\bar{a}}\partial_{\bar{a}}(K_{\bar{a}}A_{\bar{a}})]-\frac{1}{K_z}\sum_{a}\partial_{a}[\sqrt{G}G^{aa}G^{zz}F_{za}]&=&0\label{scalar_eom_2}.
\end{eqnarray}
Here we use the index $\bar{a}$ to denote $a,z$ so that $\sum_{\bar{a}}=\sum_{\bar{a}=a,z}$. Equation (\ref{rs7_eom_vector}) describes the vector modes whilst equations (\ref{scalar_eom_1}) and (\ref{scalar_eom_2}) are mixed and describe the three gauge-scalars (two physical scalar modes and one Goldstone mode). Taking suitable combinations of (\ref{scalar_eom_1}) and (\ref{scalar_eom_2}) readily gives 
\begin{eqnarray}
& &\eta^{\mu\tau}\partial_\tau\partial_\mu G_A+\xi (kz)\left[\partial_{z}\{ K_{z}\partial_{z}G_A\} +\sum_{a}\partial_{a}\{ K_{a}\partial_{a}G_A\}\right]=0\label{scalar_eom_goldstone},
\end{eqnarray}
where $G_A=(kz)\sum_{\bar{b}}\partial_{\bar{b}}(K_{\bar{b}}A_{\bar{b}})$. The decoupled scalar states described by (\ref{scalar_eom_goldstone}) are the Goldstone modes that, in the unitary gauge $\xi\rightarrow\infty$, become infinitely heavy and disappear from the spectrum. 
\subsection{KK decomposition of the vector mode\label{sec:vector_kk}}
We KK expand the vector modes $A_\mu$ as
\begin{eqnarray}
A_\mu(x^\nu,y^a,z)=\sum_{\vec{n}}A^{(\vec{n})}_\mu(x^\nu)f^{(\vec{n})}_A(z)g^{(n_a)}_+(y^a),
\end{eqnarray}
where $g^{(n_a)}_+$ is given in (\ref{ued_+_profile}). The profiles obey the orthogonality conditions
\begin{eqnarray}
\int \frac{dz}{kz}f^{(m,n_a)}_Af^{(n,n_a)}_A&=&\delta^{mn},
\end{eqnarray}
and, writing the KK masses as $m_{A,\vec{n}}$, the profiles must satisfy
\begin{eqnarray}
\left[z^2\partial_z^2-z\partial_z-\frac{m_{n_a}^2}{k^2}+m_{A,\vec{n}}^2z^2\right]f^{(\vec{n})}_A=0.\label{vector_eom_f}
\end{eqnarray}
The solution to the above is
\begin{eqnarray}
f_A^{(\vec{n})}(z)=\frac{kz}{N_A^{(\vec{n})}}\left\{J_{\nu_a}(m_{A,\vec{n}}z)+\beta_A^{(\vec{n})}Y_{\nu_a}(m_{A,\vec{n}}z)\right\},\label{vector_profile}
\end{eqnarray}
where
\begin{eqnarray}
\nu_a=\sqrt{1+\frac{m_{n_a}^2}{k^2}}.
\end{eqnarray}
The constants $\beta_A^{(\vec{n})}$ are
\begin{eqnarray}
\beta^{(\vec{n})}_A(z_*)=-\frac{J_{\nu_a-1}(m_{A,\vec{n}}z_*)}{Y_{\nu_a-1}(m_{A,\vec{n}}z_*)},
\end{eqnarray}
with the KK masses $m_{A,\vec{n}}$ following from $\beta^{(\vec{n})}_A(z_0)=\beta^{(\vec{n})}_A(z_L)$. The effective 4D action for the vector modes is thus given by
\begin{eqnarray}
\sum_{\vec{n}}\int d^4x\left\{-\frac{1}{4}\eta^{\mu\tau}\eta^{\nu\sigma}F^{(\vec{n})}_{\mu\nu}F^{(\vec{n})}_{\tau\sigma}-\frac{1}{2\xi}(\eta^{\nu\tau}\partial_\nu A^{(\vec{n})}_\tau)^2+\frac{1}{2}m_{\vec{n}}^2A^{(\vec{n})}A^{(\vec{n})}\right\}.\label{rs7_4d_vec}
\end{eqnarray}
We note that for $n_a=0$ (\ref{rs7_4d_vec}) reduces to the usual expression for a bulk vector in the RS background~\cite{Davoudiasl:1999tf} and for $n_2=0$ it reproduces the RS6 bulk vector action~\cite{Davoudiasl:2008qm}. As with the graviton KK tower the vector KK states with $n_a\ne0$ appear at TeV energies and the phenomenology of the $n_2=0$ modes will differ from the detailed study of~\cite{Davoudiasl:2008qm}.
\section{Fermions in Warped UED Models\label{sec:rs7_kk_fermi}}
We have considered already the KK spectra for gravitons, bulk vectors and the TeV brane localized Higgs. To realize a WUED scenario with the desired KK parity all SM fields must propagate in the $y^a$ space. If the SM fermions are confined to the TeV brane one would obtain some of the desirable features of UED models, however the RS approach to flavor would be lost. In this section we consider a bulk \rs fermion. We find that the appealing features of bulk RS fermions may be retained, with the 7D fermion giving rise to a single massless chiral mode which may be localized towards the Planck or TeV brane by varying a bulk fermion mass parameter. One may use this mode to model a SM fermion.

In the subsections that follow we respectively derive the equations of motion for the KK wavefunctions, present the massless mode spectrum and consider the coupling of massless fermions to gauge bosons and the Higgs.
\subsection{Bulk Fermion in RS$_{\mathbf{7}}$}
The action of the orbifold symmetries on a bulk fermion $\Psi$ is given by 
\begin{eqnarray}
Z_2'&:&\Psi(x^\mu,y^a,y)\rightarrow \Psi'(x^\mu,-y^a,y)=i P' \Gamma^5\Gamma^6 \Psi(x^\mu,y^a,y),\\
Z_2&:&\Psi(x^\mu,y^a,y)\rightarrow \tilde{\Psi}(x^\mu,y^a,-y)=i P \Gamma^7 \Psi(x^\mu,y^a,y),
\end{eqnarray}
where $\Gamma^{M}$ are the 7D gamma matrices, our conventions for which are given in Appendix~\ref{app:ferm_not}. We shall work with $P=-1$ and $P'=+1$ so that the $(Z_2',Z_2)$ parities of the components of $\Psi$ are
\begin{eqnarray}
\Psi =\left(\begin{array}{c}\psi_{-R}\ (+,-)\\\psi_{-L}\ (-,-)\\\psi_{+L}\ (+,+)\\\psi_{+R}\ (-,+)\end{array}\right),\label{7d_fermi_parity}
\end{eqnarray}
and $\psi_{+L}$ is the only field which is even under both symmetries. We note that regardless of which values are used for $P',P$ there is always only one component of $\Psi$ which is even under both $Z_2'$ and $Z_2$. The orbifold symmetries act on a Dirac mass bilinear as
\begin{eqnarray}
Z_2'&:&\overline{\Psi}\Psi\rightarrow+\overline{\Psi}\Psi,\\
Z_2&:&\overline{\Psi}\Psi\rightarrow-\overline{\Psi}\Psi,
\end{eqnarray}
so that a bulk fermion may only have a Dirac mass if the mass is odd under the action of $Z_2$, as in the RS model. 

The action for a 7D fermion\footnote{We discuss some general properties of 7D fermions in Appendix~\ref{app:ferm_not}.} in the \rs background is
\begin{eqnarray}
S_{\Psi}&=&\int d^7x\sqrt{G}\left\{\frac{i}{2}\overline{\Psi}\Gamma^{\underline{M}}e^M_{\underline{M}}\partial_M\Psi-\frac{i}{2}(\partial_M\overline{\Psi})\Gamma^{\underline{M}}e^M_{\underline{M}}\Psi-m_D\overline{\Psi}\Psi\right\},\label{7d_fermi_lagrangian}
\end{eqnarray}
where $e^M_{\underline{M}}=\mathrm{diag}(kz,kz,kz,kz,kz,1,1)$. Note that we have already dropped the spin connection terms which cancel in the above.  After rescaling the field $\Psi\rightarrow(kz)^2 \Psi$ and integrating by parts one has
\begin{eqnarray}
S_{\Psi}&=&\int d^7x\left\{i\overline{\Psi}\Gamma^{\mu}\partial_\mu\Psi+i\overline{\Psi}\Gamma^{7}\partial_7\Psi+i\frac{1}{kz}\overline{\Psi}\Gamma^{a}\partial_a\Psi-\frac{m_D}{kz}\overline{\Psi}\Psi\right\}.
\end{eqnarray}
We define the four component spinors $\psi_{+}=(\psi_{+ L},\psi_{+ R})^T$ and $\psi_{-}=(\psi_{- L},\psi_{- R})^T$ in terms of the component fields
\begin{eqnarray}
\Psi_+=(0,0,\psi_{+L},\psi_{+R})^T\quad,\quad\Psi_-=(\psi_{-R},\psi_{-L},0,0)^T
\end{eqnarray}
 and expand the action in terms of  $\psi_{\pm L,R}=P_{L,R}\psi_\pm$ as
\begin{eqnarray}
S_{\Psi}&=&\int d^7x\left\{\frac{}{}i\overline{\psi}_+\gamma^{\mu}\partial_\mu\psi_++\frac{1}{kz}\left[-\overline{\psi}_{+ L}(\partial_5-i\partial_6)\psi_{+ R}+\overline{\psi}_{+ R}(\partial_5+i\partial_6)\psi_{+ L}\right]\right.\nonumber\\
& &\qquad\quad+i\left.\overline{\psi}_-\gamma^{\mu}\partial_\mu\psi_-+\frac{1}{kz}\left[-\overline{\psi}_{- L}(\partial_5+i\partial_6)\psi_{- R}+\overline{\psi}_{- R}(\partial_5-i\partial_6)\psi_{- L}\right]\right.\nonumber\\
& &\qquad\quad+\left.\overline{\psi}_{+}(\partial_7-\frac{m_D}{kz})\psi_{-}+\overline{\psi}_{-}(-\partial_7-\frac{m_D}{kz})\psi_{+}\right\}.\nonumber
\end{eqnarray}
The KK expansions for the four component fermions are
\begin{eqnarray}
\psi_\pm(x^\mu,y^a,z)&=&\psi_{\pm L}(x^\mu,y^a,z)+\psi_{\pm R}(x^\mu,y^a,z)\nonumber\\
&=&\sum_{n,n_a}\left\{\psi^{(n,n_a)}_{ L}(x^\mu)f^{(n,n_a)}_{\pm L}(z)g^{(n_a)}_{\pm L}(y^a) +\psi^{(n,n_a)}_{ R}(x^\mu)f^{(n,n_a)}_{\pm R}(z)g^{(n_a)}_{\pm R}(y^a)\right\},\nonumber
\end{eqnarray} 
where the wavefunctions obey the following orthogonality relations,
\begin{eqnarray}
\int dz (f^{*(m,n_a)}_{+L,R}f^{(n,n_a)}_{+ L,R}+f^{*(m,n_a)}_{-L,R}f^{(n,n_a)}_{-L,R})&=&\delta^{mn},\\
\int d^2y g^{*(n_a)}_{\pm L}g^{(n_b)}_{\pm L}=\int d^2y g^{*(n_a)}_{\pm R}g^{(n_b)}_{\pm R}&=&\delta^{n_an_b}.
\end{eqnarray}
With $m_{n_a}$ given by (\ref{torus_kk_mass}) the wavefunctions $g^{(n_a)}_{\pm L,R}(y^a)$ satisfy
\begin{eqnarray}
(\partial_5\pm i\partial_6)g^{(n_a)}_{+ L,R}&=& \mp m_{n_a}g^{(n_a)}_{+R,L},\\
(\partial_5\mp i\partial_6)g^{(n_a)}_{- L,R}&=&\pm m_{n_a}g^{(n_a)}_{-R,L},
\end{eqnarray}
and may be written as
\begin{eqnarray}
g^{(n_a)}_{+ L}(y^a)&=&g^{(n_a)}_{-R}(y^a)=g^{(n_a)}_+ (y^a),\\
g^{(n_a)}_{- L}(y^a)&=&g^{(n_a)}_{+ R}(y^a)=\frac{n_1+in_2}{\sqrt{n_1^2+n_2^2}}g^{(n_a)}_- (y^a),
\end{eqnarray}
where $g^{(n_a)}_{+(-)}$ are the usual expansions for the even (odd) KK modes on the $T^2/Z_2'$ orbifold:
\begin{eqnarray}
g^{(n_a)}_+ (y^a)&=&\frac{1}{\sqrt{2}\pi R} \left(\frac{1}{\sqrt{2}}\right)^{\delta_{n_a 0}}\cos\left[\frac{n_1 y^1+n_2y^2}{R}\right],\label{ued_+_profile}\\
g^{(n_a)}_- (y^a)&=&\frac{1}{\sqrt{2}\pi R} \sin\left[\frac{n_1 y^1+n_2y^2}{R}\right].\label{ued_-_profile}
\end{eqnarray}
Finally the equations of motion for the warped direction wavefunctions are
\begin{eqnarray}
\left[-\partial_7-\frac{c}{z}\right]f^{(n,n_a)}_{+ R}+\frac{c_{n_a}}{z}f^{(n,n_a)}_{- R}&=&-m_{n,n_a}f^{(n,n_a)}_{- L},\label{rs7_vecfermi_eom_2a}\\
\left[\partial_7-\frac{c}{z}\right]f^{(n,n_a)}_{- R}-\frac{c_{n_a}}{z}f^{(n,n_a)}_{+ R}&=&-m_{n,n_a}f^{(n,n_a)}_{+ L},\label{rs7_vecfermi_eom_2b}\\
\left[-\partial_7-\frac{c}{z}\right]f^{(n,n_a)}_{+L}+\frac{c_{n_a}}{z}f^{(n,n_a)}_{-L}&=&-m_{n,n_a}f^{(n,n_a)}_{-R},\label{rs7_vecfermi_eom_2c}\\
\left[\partial_7-\frac{c}{z}\right]f^{(n,n_a)}_{-L}-\frac{c_{n_a}}{z}f^{(n,n_a)}_{+L}&=&-m_{n,n_a}f^{(n,n_a)}_{+R},\label{rs7_vecfermi_eom_2d}
\end{eqnarray}
where we have defined the dimensionless masses
\begin{eqnarray}
c=\frac{m_D}{k}\quad,\quad c_{n_a}=\frac{m_{n_a}}{k}.
\end{eqnarray}
Noting the parities (\ref{7d_fermi_parity}) one may use the equations of motion to obtain the boundary conditions,
\begin{eqnarray}
\left.f^{(n,n_a)}_{- L,R}\right|_{z_*}&=&0,\label{fermi_bc_odd}\\
\left.\left(\partial_z+\frac{c}{z}\right)f^{(n,n_a)}_{+L, R}\right|_{z_*}&=&0.\label{fermi_bc_even}
\end{eqnarray}
\subsection{Massless Fermion Modes\label{sec:mssless_fermi}}
Our primary interest is in the spectrum of massless modes as this will determine the viability of employing a bulk 7D fermion. Setting $m_{n,n_a}=0$ in (\ref{rs7_vecfermi_eom_2a})-(\ref{rs7_vecfermi_eom_2d}) one finds that the profiles of the massless modes must satisfy the following differential equation,
\begin{eqnarray}
(z^2\partial_z^2+z\partial_z-c^2-c_{n_a}^2)f^{(n,n_a)}_{\pm L,R}&=&0,
\end{eqnarray}
which has solutions of the general form
\begin{eqnarray}
 f^{(n,n_a)}_{\pm L,R}=A_{\pm L,R}^{(n,n_a)}z^{c_-}+B_{\pm L,R}^{(n,n_a)}z^{c_+},
\end{eqnarray}
where $c_{\pm}=\pm\sqrt{c^2+c^2_{n_a}}$ and $A,B$ are constants. The boundary conditions (\ref{fermi_bc_odd}) are only satisfied for $c_{n_a}=n_a=0$ and the single massless mode in the spectrum is given by $f_{+L}^{(0,0)}$ with wavefunction
\begin{eqnarray}
f_{+L}^{(0,0)}(z)\propto z^{-c}.\label{fermi_zero}
\end{eqnarray}
This is the usual form for a chiral zero mode in the RS background and thus, as per RS, one may localize the zero mode towards the UV or IR brane by changing the value of $c$ (or equivalently the bulk Dirac mass $m_D$). Equation (\ref{fermi_zero}) is our main result for this section. It tells us that, starting with a massive 7D fermion, one obtains a single massless chiral mode in the effective 4D theory which may be identified with a SM fermion. This motivates the study of \rs models with bulk SM fermions and, more generally, the construction of WUED models on the \rs background. As the zero mode profile (\ref{fermi_zero}) is the same as that obtained in RS models one expects much of the flavor structures studied already in RS models to go through in the present framework.

For completeness, the normalized wavefunction for the massless chiral mode is,
\begin{eqnarray}
(kz)^2f_{+L}^{(0,0)}(z)=\sqrt{\frac{k(1/2-c)}{(kz_L)^{(1-2c)}-1}}(kz)^{2-c},
\end{eqnarray}
where we retain the factor of $(kz)^2$ previously scaled out. This is identical to the usual RS profile~\cite{Grossman:1999ra,Gherghetta:2000qt}. We shall not solve equations (\ref{rs7_vecfermi_eom_2a})-(\ref{rs7_vecfermi_eom_2d}) for the general case but rather content ourselves with providing the solutions for the special cases of $n_a=0$ and $m_D=0$ in Appendix~\ref{app:fermi_kk}.
\subsection{Flavor and Gauge Couplings\label{sec:flavor_gauge}}
We now obtain the Yukawa and gauge couplings for a chiral zero mode fermion. The RS approach to flavor carries through to \rs, with the Yukawa coupling between two bulk fermions $\Psi_{1,2}$ and a TeV brane localized scalar given by
 \begin{eqnarray}
S_{Yuk}&=&-\frac{\lambda_Y}{k^2} \int d^7x \sqrt{\bar{G}}\tilde{H}\Psi_1\Psi_2\delta(z-z_L)\nonumber\\
&=&-\sum_{n,m, n_a}\frac{\lambda_{eff}^{n,m,n_a}}{k}\int d^4x \langle H^{(0)}\rangle\psi^{(n,n_a)}_1\psi^{(m,n_a)}_2 + ....\label{yuk_rs7}
\end{eqnarray}
Here the dots denote terms involving KK excitations of the scalar, $\lambda_Y$ is dimensionless and $H= (kz_L)^{-1} \tilde{H}$ is the canonically normalized scalar. The effective coupling is defined as
\begin{eqnarray}
\left.\lambda_{eff}^{n,m,n_a}=\frac{\lambda_Y}{e^{-k\pi r_c}k}f^{(n,n_a)}_1f^{(m,n_a)}_2\right|_{z=z_L},
\end{eqnarray}
and we write the Higgs VEV as
\begin{eqnarray}
\langle H^{(0)}\rangle =e^{-k\pi r_c}v_0 \equiv  \frac{v}{2\pi R},
\end{eqnarray}
where $v= 2\pi Re^{-k\pi r_c}v_0$ is the weak scale and we note that $[v_0]=2$. The effective 4D fermion mass induced by (\ref{yuk_rs7}) is 
\begin{eqnarray}
m_\psi^{n,m,n_a}= \lambda_{eff}^{n,m,n_a}\frac{v}{2\pi kR},
\end{eqnarray}
 and we are interested primarily in the mass of the zero mode fermions, which have the quantum numbers $n,m,n_a=0$. One observes that the zero mode masses are the same as those obtained in RS models, modulo the factor $(2\pi kR)^{-1}$, which is $\sim6$ for $kR\sim1$.

The coupling between a bulk fermion and a bulk gauge boson in \rs is
\begin{eqnarray}
S_{\Psi,A}&=&g_7\int d^7x\sqrt{G}e^M_{\underline{M}}\overline{\Psi}\Gamma^{\underline{M}}\Psi A_M\nonumber\\
&=&g_7\int d^7x \left\{\overline{\psi}_+\gamma^{\mu}\psi_++\overline{\psi}_-\gamma^{\mu}\psi_-\right\}A_\mu +...\nonumber\\
&=&g_{eff}^{n}\int d^4x\bar{\psi}_L^{(0,0)}\gamma^{\mu}\psi^{(0,0)}_L A_\mu^{(n,0)}+.... ,\label{7d_fermi_gauge}
\end{eqnarray}
where in the last line we have retained only the terms with the chiral mode and defined the coupling
\begin{eqnarray}
 g_{eff}^{n}=g_4\sqrt{2\pi r_c}\int dz f^{*(0,0)}_{+L}f_{+L}^{(0,0)}f_A^{(n,0)}.\label{RS_like_coupling_ffa}
\end{eqnarray}
Here we write the couping in terms of the 4D coupling of the massless vector mode $g_4=g_{eff}^{0}$, which is identified with a SM gauge coupling $g_4=g_{SM}$. The vector modes $A_\mu^{(n,0)}$ are the usual RS KK modes and as expected the KK parity has not altered the coupling between SM fermions and these modes. The vector modes $A_\mu^{(n,n_a)}$ with $n_a\ne0$ on the other hand do not couple directly to the fermion zero mode current $J^\mu_{SM}\sim \bar{\psi}_L^{(0,0)}\gamma^{\mu}\psi^{(0,0)}_L$. These modes correspond to excitations on $T^2$ and are the UED-like modes. As in UED models two SM fermions cannot annihilate directly into these states.
\section{KK Parity and the LKP\label{sec:dm_kk}}
One of the the most appealing features of UED models is the existence of a KK parity which plays two important roles. Firstly, because the lightest KK modes are odd under the KK parity they must be pair produced. This suppresses their contribution to precision observables, thereby weakening the bounds on the KK scale and providing a greater likelihood that the KK scale will be experimentally accessible\footnote{The suppression of new physics effects induced by the KK parity can also make the detection of the KK modes more challenging for precisely the same reason. In fact it may be easier to experimentally observe the second level KK states~\cite{Datta:2005zs}, which may also play a role in cosmology~\cite{Kakizaki:2005en}}. Secondly the KK parity requires the lightest KK odd state to be stable and thus provides an appealing DM candidate~\cite{Servant:2002aq}.

The usual arguments leading to the existence of a KK parity in UED models hold in the \rs background. The localized brane tensions necessarily break $SO(1,6)$ so that the full 7D Lorentz symmetry does not hold. Before compactification of the $x^{5,6}$ directions the \rs space is $AdS_5\times \mathcal{R}^2$. The 6D Lorentz symmetry $SO(1,5)\subset SO(1,6)$ is also broken because the bulk and brane sources (\ref{rs7_cc}) and (\ref{rs7_tension}) are not homogeneous. However the subgroup $SO(1,3)\times U(1)_{56}$, where $U(1)_{56}$ is the group of rotations in the $x^{5,6}$ plane, is preserved so that prior to compactification the reduced Lorentz symmetry $SO(1,3)\times U(1)_{56}$ holds. Compactification of $x^{5,6}$ via $\mathcal{R}^2\rightarrow T^2/Z_2'$ breaks $U(1)_{56}$ but a discrete subgroup  $Z_2^{KK}$ remains unbroken. Provided that all SM fields propagate in the $x^{5,6}$ directions the KK parity $Z_2^{KK}$ remains a good symmetry of the theory. The $Z_2^{KK}$ parity of a field is given by $(-1)^{n_1+n_2}$ so all modes for which $n_1+n_2$ is an odd integer cannot be individually produced. The lightest such state is an absolutely stable DM candidate~\cite{Servant:2002aq}.

Let us say a few more words about the presence of the $Z^{KK}_2\subset U(1)_{56}$ symmetry. As in UED models, the preservation of this symmetry  necessitates an identification of the physics at related $T^2/Z_2'$ fixed points. Without this identification localized operators would in general break the KK parity. However loop induced localized operators automatically satisfy this constraint so that it is reasonable to expect the KK parity to remain as a good symmetry. The existence of the KK parity in the low energy theory also requires a method of radius stabilization which does not break $Z^{KK}_2$. This demand is common to all UED models and strictly speaking one cannot claim to have completely motivated the existence of a stable DM candidate in a UED or a WUED model unless a method of radius stabilization which preserves the KK parity is incorporated. We have nothing to add to this important issue here but remind the reader that ultimately it should be addressed in both the UED and WUED framework.

In RS models the KK modes of bulk fields do not possess a conserved KK parity because the warping of the extra dimension breaks the translational symmetry along the extra dimension in a maximal fashion. Consequently the RS KK modes may be individually produced and their contribution to precision observables in the simplest constructs produces strong bounds on the KK scale (though additional model building allows one to lower most of these bounds, for a sample see~\cite{Agashe:2003zs}). In \rs the usual RS modes correspond to the $n_a=0$ modes of a bulk field. These have even KK parity so that KK parity does not effect the production and stability of these states and the usual bounds obtained in RS models will in general apply. However the important observation here is that the production of the extra UED KK modes with $n_a\ne0$ will be suppressed. 

Given that the WUED framework admits a stable LKP one would like to know if this particle may  be a suitable DM candidate. Detailed calculations of relic abundances are beyond the scope of this work, but we offer the following comments regarding the LKP. As in UED models all of the SM fields will have KK partners and of these the likely candidates for a suitable LKP include the KK graviton, photon, neutral scalar and right-handed neutrino. The LKP will have the quantum numbers $n_1^2+n_2^2=1$ and we show the tree level mass of the lightest KK modes with $(n_1,n_2)=(1,0)$ for the graviton ($h$), vector ($A$), fermion ($\Psi$) and the CP odd neutral scalar ($A^0$) in Table~\ref{table:masses}. The masses are given in units of $e^{-k\pi r_c}k=1$~TeV and the fermion bulk mass has been set to zero ($m_D=0$) so the fermion KK masses are found using (\ref{massless_fermi_kkmass}). One observes immediately that the gauge KK masses are generically lighter than the graviton KK masses, whilst the fermion and scalar KK modes are lighter again. It seems that the likely candidate for the LKP is $A^0$, however this conclusion may change after radiative corrections to the KK masses are included~\cite{Cheng:2002iz}. The scalar KK masses receive radiative corrections that are sensitive to the cutoff and although the brane cutoff is warped down, the corrected scalar masses could lie anywhere between the tree value and the order TeV cutoff\footnote{See~\cite{Casagrande:2008hr}~(\cite{Dobrescu:2007xf}) for a discussion of radiative corrections to the Higgs mass (KK Higgs mass) in RS (UED) models.}. The fermion KK masses also receive radiative corrections and turning on a non-zero bulk mass will further modify the masses. Whilst the tree level masses alone do not definitively determine the LKP, it is possible that the KK partner of the right-chiral neutrino or the neutral Goldstone mode may be the lightest\footnote{Note that electroweak symmetry induces mixing between the Higgs KK modes and the gauge-scalars such that the Goldstones of the KK vectors and the physical scalars are linear combinations of these states~\cite{DeCurtis:2002nd}.}. The KK partner of the neutral Goldstone, $A^0$, is an interesting candidate as it is already known that the presence of a brane localized scalar with an order TeV mass in RS models can provide a viable DM candidate~\cite{RP_RSDM}.
\begin{table}[ht]
\centering
\begin{tabular}{|l||l|l|l|l|}
\hline
$kR$&1&2&3&4\\ \hline
$m_h$&4.14&3.91&3.87&3.85 \\
$m_A$&3.02&2.58&2.49&2.46\\
$m_\Psi$&2.04&1.43&1.48&1.49\\
$m_{A^0}$&1&0.5&0.33&0.25\\
\hline
\end{tabular}
\caption{Tree level masses of the lightest KK modes with $(n_1,n_2)=(1,0)$ for the graviton ($h$), vector ($A$), fermion ($\Psi$) and CP odd neutral scalar ($A^0$). The fermion has no bulk mass ($m_D=0$) and the masses are given in units of $e^{-k\pi r_c}k=1$~TeV.}
\label{table:masses}
\end{table}

It may be that the LKP is not actually the KK partner of a SM field. The 7D space admits additional particles in the form of metric fluctuations and gauge boson polarizations along the compact dimensions. As such the LKP may be a gauge-scalar mode, as can occur in 6D UED models~\cite{Dobrescu:2007ec}. The radiative corrections to the KK masses of a non-Abelian gauge-scalar~\cite{Dobrescu:2007xf} are typically dominated by negative contributions from fermion loops. This reduces the gauge-scalar mass relative to the KK scale and can result in the SM hypercharge gauge-scalar being the LKP~\cite{Dobrescu:2007ec}. It is an interesting problem to find the KK spectrum for the gauge-scalars in the present framework and to determine if a gauge-scalar LKP may be a suitable DM particle. 

We note that as one increases $kR$ beyond the values in Table~\ref{table:masses} the KK mass of the $(n_1,n_2)=(1,0)$ modes of the graviton (photon) asymptote to $3.83$ ($2.44$).  This is the mass of the first KK graviton (photon) in RS and thus one cannot make the $(1,0)$ modes lighter than the $n=1$ modes in the corresponding RS KK towers. The KK partners of the SM Goldstones on the other hand decrease without lower bound as $kR$ increases and the lightest tree level masses for these KK scalars exceeds $10^2$~GeV only for $kR<10$.
\section{Standard Model on the TeV Brane\label{sec:sm_tev}}
Before concluding we would like to comment on an alternative possibility in the \rs geometry. To date we have assumed that only the SM Higgs is confined to the TeV brane, with all SM fields propagating in the extra toroidal spatial dimensions. This is necessary if one seeks to retain the RS approach to flavor. If one abandons RS flavor and accepts instead a trivial flavor structure (as in UED models) one can place all the SM fields on the TeV brane and retain the UED structure. The resulting model is essentially a 6D UED model with additional graviton KK states from the warped dimension. A similar framework has already been investigated in~\cite{Appelquist:2002ft} with all SM fields localized on the 6D TeV brane of a slice of $AdS_7$. There are some important differences between realizing a UED scenario by placing the SM on the 6D TeV brane of \rs verses the 6D TeV brane of a slice of $AdS_7$. One main difference results from the way in which the warp factor couples to brane localized matter in the two spaces. To observe this difference consider a brane localized real scalar in 7D with action, 
\begin{eqnarray}
S_\Phi&=&\frac{1}{2}\int d^{7}x\sqrt{\bar{G}}\left\{G^{\bar{M}\bar{N}}D_{\bar{M}}\Phi D_{\bar{N}}\Phi -m_{\Phi}^2\Phi^2\right\}\delta(z-z_L),
\end{eqnarray}
where the metric may be the brane restriction of the \rs metric or the $AdS_7$ metric, whose form we take as
\begin{eqnarray}
ds^2_{AdS_7}&=&\frac{1}{(kz)^2}\left[\eta_{\mu\nu}dx^\mu dx^\nu-\delta_{ab}dy^ady^b-dz^2\right].\label{ads7_conformal_metric}
\end{eqnarray}
The KK masses of the resulting tower in the 4D theory are
\begin{eqnarray}
m_{\Phi,n_a}^2=\left\{\begin{array}{ll}m_{\Phi}^2e^{-2k\pi r_c}+m_{n_a}^2e^{-2k\pi r_c}&\mathrm{for}\quad RS_7,\\ m_{\Phi}^2e^{-2k\pi r_c}+m_{n_a}^2&\mathrm{for}\quad AdS_7.\end{array}\right.\label{scalar_kk}
\end{eqnarray}
Observe that in both cases the bare mass $m_\Phi$ is warped down but \emph{only} in \rs is the KK scale of the brane scalar also warped down. Thus if the SM is localized on the TeV brane the weak scale can be realized via the warping of order $M_*$ input parameters in both \rs and $AdS_7$. However only in \rs is the weak scale connected to the UED KK scale with the KK modes of SM fields expected at $e^{-k\pi r_c}R^{-1}\sim$~TeV. In $AdS_7$ the KK masses are of order $R^{-1}$ and only appear at the weak scale if $R^{-1}$ is independently taken to be at the TeV scale. 

Note that the above difference is readily understood. Recall that the RS metric is conformally flat. After integrating out the extra dimension in RS models one must rescale the wavefunction for any brane localized fields to bring the kinetic terms to a canonical form. The Lagrangian of any brane localized field is invariant under this transformation provided it is classically scale invariant\footnote{More strictly speaking scale invariance is a good classical symmetry which is however broken quantum mechanically. This can have interesting effects for brane fields in RS models; see~\cite{Dienes:1999sz}.}. However if the brane Lagrangian contains dimensionful parameters this rescaling is not a good symmetry and the warp factor is manifest in the effective 4D Lagrangian. This is exactly what occurs in the RS model and is the reason why the RS model provides a candidate solution to the hierarchy problem; the scale invariance of the brane Lagrangian is broken by both the Higgs mass and the cutoff so that in the effective 4D theory both the bare Higgs mass \emph{and} the loop corrections to the Higgs mass, which depend on the cutoff, are warped down to the Tev scale.

The $AdS_7$ metric is also conformally flat so that, as in RS models, only parameters in the TeV brane Lagrangian which explicitly break scale invariance (like the scalar mass) are warped down. The \rs metric on the other hand is not conformally flat for $R\ne 0$ so that the warp factor couples to all brane parameters which break the 4D scale invariance. Both kinetic energy along the toroidal directions and the explicit scalar mass break scale invariance so that both the bare and KK masses are warped down to the TeV scale in the effective theory.

Note that this observation holds more generally for spacetimes of the form $AdS_5\times \mathcal{M}^\delta$, where $\mathcal{M}^\delta$ is a compact space with $\delta$ dimensions of radius $R$. If the Lagrangian for a UED model with $\delta$ compact extra dimensions is localized at the TeV brane of the space $AdS_5\times \mathcal{M}^\delta$ the KK masses of the SM fields will be warped down to $\sim$~TeV energies, as is usually assumed in the UED literature. Thus in addition to the embedding of the 6D UED model in \rs discussed above one could consider, for example, the 5D UED model embedded in RS6. The generic experimental signature for such an extension of the UED framework would be the observation of warped KK gravitons in addition to the TeV scale KK modes of SM fields.
\section{Conclusion\label{sec:conc}}
We have developed the necessary ingredients for the construction of WUED models based on the geometry $AdS_5\times T^2/Z_2$. The KK spectra for gravitons, bulk vectors and the TeV brane localized Higgs boson were derived and we have shown that, starting with a massive 7D fermion, one may obtain a single chiral massless mode whose profile is readily localized towards the Planck or TeV brane. This allows one to consider models of flavor as in RS. Our solution also admits the familiar KK parity of UED models so that the KK odd states cannot be singly produced and the lightest such state is a stable DM candidate. As an additional feature the $AdS_5$ warping ensures that the excited modes on the torus (including the DM candidate) appear at TeV energies, as is usually assumed in UED models. This is true even though the Planck scale sets the dimensions for the torus. Finally we have noted that the observed connection between the weak scale and the UED KK scale persists more generally in spacetimes of the form $AdS_5\times \mathcal{M}^\delta$ when one abandons the RS approach to flavor and localizes the UED Lagrangian on the TeV brane.

There are a number of obvious directions for further work here. Of particular interest is the viability of the LKP as a DM candidate and a determination of the gauge-scalar KK spectra. 
\section*{Acknowledgments}
The author thanks B.~Batell, H.~Davoudiasl, B.~Dobrescu and T.~Rizzo. This work was supported in part by the Natural Science and Engineering Research Council of Canada.
\appendix
\section*{Appendix}
\section{Generalization to $\mathbf{AdS_5\times T^n}$\label{app:rsd}}
The \rs solution is readily generalized to RS$_d$ for $d>7$. The metric remains as in (\ref{rs7_metric}) however now the label for the extra dimensions $a$ takes the values $a=1,2,..,n$. The sources generalize to
\begin{eqnarray}
\Lambda^M_N=\left(\begin{array}{ccc} \Lambda I_{4}& 0&0\\0& \bar{\Lambda}I_{n}&0\\0&0&\Lambda\end{array}\right)\quad,\quad T^{\bar{M}}_{h,v\bar{N}}=\left(\begin{array}{cc}V_{h,v} I_{4}& 0\\0&\bar{V}_{h,v}I_{ n}\end{array}\right),\label{rsd_tension}
\end{eqnarray}
where $I_{n}$ denotes the $n$ dimensional identity matrix. The solution to the Einstein equations is
\begin{eqnarray}
\sigma&=&\sqrt{\frac{-\Lambda}{24M_*^{3+n}}}|y|\equiv k|y|,\label{rsd_sigma_sol}
\end{eqnarray}
with
\begin{eqnarray}
V_h=-V_v=24kM^{3+n}_*\quad,\quad V_{h,v}=\frac{3}{4}\bar{V}_{h,v}\quad,\quad\Lambda=\frac{3}{5}\bar{\Lambda},
\end{eqnarray}
and the 4D Planck mass is
\begin{eqnarray}
M_{Pl}^2=(2\pi R)^n \frac{M^{3+n}_*}{k}[1-e^{-2k\pi r_c}].\label{rsd_planck}
\end{eqnarray}
\section{Fermions in 7D\label{app:ferm_not}}
The generators of the 7D Lorentz group $SO(1,6)$ for the spin 1/2 representation are
\begin{eqnarray}
S^{MN}=\frac{\Sigma^{MN}}{2}=\frac{i}{4}[\Gamma^M,\Gamma^N],
\end{eqnarray}
where the 7D gamma matrices satisfy
\begin{eqnarray}
\{\Gamma^M,\Gamma^N\}=2\eta^{MN}I,\label{clifford_algebra}
\end{eqnarray}
and $\eta^{MN}=\mathrm{diag}(1,-1,-1,...)$. The minimum dimensionality of the seven matrices satisfying the Clifford algebra (\ref{clifford_algebra}) is $8\times8$ so that fermions are described by spinors with eight components. Our representation of the $\Gamma$-matrices is as follows. For $M=0,1,2,3,5,6$ we use
\begin{eqnarray}
\Gamma^M=\left(\begin{array}{cc}0&\Sigma^M\\\bar{\Sigma}^M&0\end{array}\right)\label{6D_gamma},
\end{eqnarray}
where 
\begin{eqnarray}
\Sigma^0&=&\bar{\Sigma}^0=\gamma^0\gamma^0\quad,\quad\Sigma^i=-\bar{\Sigma}^i=\gamma^0\gamma^i\\
\Sigma^5&=&-\bar{\Sigma}^5=i\gamma^0\gamma^5\quad,\quad \Sigma^6=-\bar{\Sigma}^6=\gamma^0,
\end{eqnarray}
and for definiteness we employ the Weyl representation of the Dirac gamma matrices
\begin{eqnarray}
\gamma^0&=&\left(\begin{array}{cc}0&1\\1&0\end{array}\right)\quad,\quad\gamma^i=\left(\begin{array}{cc}0&\sigma^i\\-\sigma^i&0\end{array}\right)\quad,\quad\gamma^5=\left(\begin{array}{cc}-1&0\\0&1\end{array}\right).\label{dirac_matrices_weyl}
\end{eqnarray}
In 4D the projection operators $P_{R,L}=\frac{1}{2}(1\pm\gamma^5)$ project out the right- and left-chiral components of a Dirac spinor. These operators may be generalized to 7D as
\begin{eqnarray}
 P^7_{R,L}&=&\frac{1}{2}(1\pm i\Gamma^0\Gamma^1\Gamma^2\Gamma^3).
\end{eqnarray}
The final gamma matrix is
\begin{eqnarray}
\Gamma^7&=&i\bar{\Gamma}\equiv i\Gamma^0\Gamma^1\Gamma^2\Gamma^3\Gamma^5\Gamma^6=i\left(\begin{array}{cc}-I&0\\0&I\end{array}\right),
\end{eqnarray}
which may be used to define the projection operators
\begin{eqnarray}
 P_\pm&=&\frac{1}{2}(1\pm\bar{\Gamma}).
\end{eqnarray}
Thus one may label the components of the 7D spinor with their 6D chirality ($\pm$) and their 4D chirality ($R,L$) as
\begin{eqnarray}
\Psi =\left(\psi_{-R},\psi_{-L},\psi_{+L},\psi_{+R}\right)^T,\label{7d_fermi}
\end{eqnarray}
and one may define $\Psi_{\pm}=P_{\pm}\Psi$. The 7D Dirac conjugate field is given by $\overline{\Psi}=\Psi^\dagger\Gamma^0$.
\section{Fermion KK profiles: Some Special Cases\label{app:fermi_kk}}
\subsection{Zero UED Momentum}
For $n_a=0$ equations (\ref{rs7_vecfermi_eom_2a})-(\ref{rs7_vecfermi_eom_2d}) may be separated as
\begin{eqnarray}
(z^2\partial_z^2-(c^2\pm c)+m_{n,0}^2z^2)f^{(n,0)}_{\pm L,R}&=&0,
\end{eqnarray}
which have solutions
\begin{eqnarray}
f_{\pm L,R}^{(n,0)}(z)=\frac{\sqrt{kz}}{N_{\pm L,R}^{(n,0)}}\left\{J_{\alpha_\pm}(m_{n,0}z)+\beta_{\pm L,R}^{(n,0)}Y_{\alpha_{\pm}}(m_{n,0}z)\right\},\label{fermi_kk_na_zero}
\end{eqnarray}
with $\alpha_{\pm}=|c\pm \frac{1}{2}|$. The boundary conditions, combined with Bessel function identities, give
\begin{eqnarray}
\beta_{-L,R}^{(n,0)}(z_*)=\beta_{+L,R}^{(n,0)}(z_*)=-\frac{J_{\alpha_{+}}(m_{n,0}z_*)}{Y_{\alpha_+}(m_{n,0}z_*)},
\end{eqnarray}
and the KK masses result from $\beta_{\psi}^{(n,0)}(z_0)=\beta_{\psi}^{(n,0)}(z_L)$. The equations of motion require $N_{+L,R}^{(n,0)}=N_{-R,L}^{(n,0)}\equiv N_{\Psi}^{(n,0)}$ so one may write the solutions as
\begin{eqnarray}
f_{\pm L,R}^{(n,0)}(z)=\frac{\sqrt{kz}}{N_{\Psi}^{(n,0)}}\left\{J_{\alpha_{\pm}}(m_{n,0}z)+\beta_{\Psi}^{(n,0)}Y_{\alpha_{\pm}}(m_{n,0}z)\right\}.\label{fermi_kk_na_zero'}
\end{eqnarray}
\subsection{Vanishing Bulk Mass}
Setting $m_D=0$ one finds that the wavefunctions satisfy the coupled equations
\begin{eqnarray}
(z^2\partial_z^2-c_{n_a}^2+m_{n,n_a}z^2)f^{(n,n_a)}_{\pm L,R}+c_{n_a}f^{(n,n_a)}_{\mp L,R}&=&0,
\end{eqnarray}
which gives
\begin{eqnarray}
f_{\pm L,R}^{(n,n_a)}(z)=\frac{1}{\sqrt{2}}\frac{\sqrt{kz}}{N_{\Psi}^{(n,n_a)}}\left\{J_{\nu_-}\pm J_{\nu_+}+\beta_{\Psi}^{(n,n_a)}\left[Y_{\nu_-}\pm Y_{\nu_+}\right]\right\}.\label{fermi_kk_mD_zero'}
\end{eqnarray}
We have suppressed the argument of the Bessel functions, which is $(m_{n,n_a}z)$, and written the order of the Bessels as $\nu_{\pm}=|c_{n_a} \pm \frac{1}{2}|$. As per usual the masses follow from  $\beta_{\psi}^{(n,n_a)}(z_0)=\beta_{\psi}^{(n,n_a)}(z_L)$ and the boundary conditions give
\begin{eqnarray}
\beta_{\Psi}^{(n,n_a)}(z_*)&=&-\frac{J_{\nu_-}(m_{n,n_a}z_*)-J_{\nu_+}(m_{n,n_a}z_*)}{Y_{\nu_-}(m_{n,n_a}z_*)-Y_{\nu_+}(m_{n,n_a}z_*)}.\label{massless_fermi_kkmass}
\end{eqnarray}

\end{document}